**Title**

- Cloud-aerosol interactions in subtropical marine stratocumulus weaken in a warmer climate.
- Aerosol-Cloud Interactions under Climate Change.

**Authors**

Hongwei Sun[1]*, Peter N. Blossey[1], Robert Wood[1], Ehsan Erfani[2], Sarah Doherty[1,3], Je-Yun Chun[1]

**Affiliations**

[1] Department of Atmospheric and Climate Science, University of Washington, Seattle, WA, USA

[2] Division of Atmospheric Sciences, Desert Research Institute, Reno, NV, USA

[3] Cooperative Institute for Climate, Ocean, & Ecosystem Studies, University of Washington, Seattle, WA, USA

*Corresponding author: Hongwei Sun (hongwei8@hawaii.edu)

**Abstract**

Radiative effects of aerosol-cloud interactions constitute the most uncertain climate forcing of the Earth system, making it important to understand how they may change with climate. We conduct 3-day-long large-eddy simulations of a stratocumulus-to-cumulus transition along an airmass-following trajectory over the Northeast Pacific Ocean. By perturbing boundary-layer aerosol concentrations, we simulate aerosol-cloud interactions in both present-day and doubled-$CO_2$ conditions. Aerosol-induced cloud changes, including the Twomey effect and adjustments of cloud fraction and liquid water path, are inhibited in a doubled-$CO_2$ climate. Decomposing the aerosol-induced cloud radiative effect change (ΔCRE) reveals that aerosol-induced cloud fraction changes dominate ΔCRE. Overall, doubling $CO_2$ attenuates aerosol-induced ΔCRE (i.e., cooling) by >30% in our simulations. Our results also show that low cloud feedbacks are sensitive to the background aerosol concentration, highlighting the interplay between climate forcings and feedbacks. These results may aid in predicting the cooling potential of marine cloud brightening in a changing climate.

**Teaser**

Doubling $CO_2$ inhibits aerosol-cloud interactions, weakening aerosol-induced cloud radiative effects from marine cloud brightening.

## MAIN TEXT

## Introduction

Clouds are one of the most important components in our climate system and have significant influences on, for example, radiation, precipitation, circulation, and temperature (*1-3*). Aerosols (both natural and anthropogenic) can serve as cloud condensation nuclei (CCN) to contribute to cloud formation (*4, 5*), and then further influence cloud radiative effects (e.g., Twomey effects (*6, 7*)). A better understanding of aerosol-cloud interactions would help improve the accuracy of weather and climate predictions. For example, recent studies show that reduced aerosols due to the International Maritime Organization 2020 (IMO2020) regulation, which reduced commercial shipping fuel sulfur emissions, could potentially increase global warming by dimming the marine low clouds (*8-12*).

It is known that aerosols, clouds, and their interactions introduce huge uncertainties in climate forcing in global climate model (GCM) simulations (*13-16*). The inability of GCMs to resolve subgrid processes (e.g., boundary layer turbulence and moist convection) makes them highly dependent on various parameterizations (*17, 18*). On the other hand, large-eddy simulation (LES) has a sufficiently high vertical and horizontal resolution to explicitly simulate much of the energy-containing turbulence that affects clouds and is a widely used tool to simulate cloud microphysics (*19, 20*) and cloud organization (*21-23*), though a limitation is that these simulations often have modest domain sizes (up to ~100 km). There are many LES studies focusing on how climate change influences stratocumulus clouds, such as the CFMIP/GASS Intercomparison of Large-Eddy and Single-Column Models (CGILS) (*24-26*). However, the influences of global warming on aerosol-cloud interactions in the marine boundary layer (MBL) have not been well investigated in LES studies.

A distinctive feature of marine low clouds over eastern ocean basins is the stratocumulus-to-cumulus transition (SCT) (*27-33*), a cloud regime shift that occurs as lower-tropospheric air masses from stratocumulus-dominated regions are advected equatorward by the trade winds. This study employs a Lagrangian LES method (*34, 35*) to simulate SCT along an airmass-following trajectory using the SAM LES model (*8*). We carry out simulations with and without perturbations to MBL aerosol concentrations (control vs. perturbed cases) in both present-day conditions (PD) and a warmer climate. The warmer climate scenario (P2CO2) features doubled atmospheric $CO_2$ and a 2 K increase in sea surface temperature (SST) and was derived following the approach in a previous study (*36*), henceforth BB2014, which is also the source of the P4CO2 scenario. Table 1 summarizes all the simulations and additional model configuration details are provided in the Methods section.

**Table 1. Summary of all simulations.**

| Simulation | Climate scenario | Initial MBL aerosols | Aerosol perturbations | Simulation duration |
|---|---|---|---|---|

| PD_L$^{c}$ | Present day (PD) | MBL_30 (Low) | No (control case) | 3 days |
|---|---|---|---|---|
| PD_L$^{p}$ | Present day | MBL_30 | Yes (perturbed case) | 3 days |
| P2CO2_L$^{c}$ | 2x$CO_2$ climate (P2CO2) | MBL_30 | No | 3 days |
| P2CO2_L$^{p}$ | 2x$CO_2$ climate | MBL_30 | Yes | 3 days |
| P4CO2_L$^{c}$ | 4x$CO_2$ climate (P4CO2) | MBL_30 | No | 1 day |
| P4CO2_L$^{p}$ | 4x$CO_2$ climate | MBL_30 | Yes | 1 day |
| PD_H$^{c}$ | Present day | MBL_100 (High) | No | 1 day |
| PD_H$^{p}$ | Present day | MBL_100 | Yes | 1 day |
| P2CO2_H$^{c}$ | 2x$CO_2$ climate | MBL_100 | No | 1 day |
| P2CO2_H$^{p}$ | 2x$CO_2$ climate | MBL_100 | Yes | 1 day |
| P4CO2_H$^{c}$ | 4x$CO_2$ climate | MBL_100 | No | 1 day |
| P4CO2_H$^{p}$ | 4x$CO_2$ climate | MBL_100 | Yes | 1 day |

Note: Most analyses in this study use the first four simulations highlighted in the table (i.e., PD_L$^{c}$, PD_L$^{p}$, P2CO2_L$^{c}$, P2CO2_L$^{c}$). MBL_30 refers to the MBL with a lower initial aerosol concentration (MBL-average) of 30 $cm^{-3}$. MBL_100 refers to the MBL with a higher initial aerosol concentration of 100 $cm^{-3}$.

## Results

### Aerosol-cloud interactions

In the control cases without aerosol perturbation, aerosol number concentration (Na) stabilizes around 30 $cm^{-3}$ (blue and red solid lines in Figure 1b), averaged within the MBL. The aerosol concentration is slightly higher in the P2CO2 than in the PD, which may result from less interstitial scavenging in a doubled-$CO_2$ climate (Figure S2.a). Due to the lack of cloud condensation nuclei (CCN) (solid lines in Figure 1b) in the MBL, clouds in the control cases quickly drizzle out (solid lines in Figures 1c and 1f), resulting in a low cloud fraction (solid lines in Figure 1h). Note that the overcast stratocumulus cloud breaks up much earlier in the PD_L$^{c}$ simulation than in BB2014, likely because BB2014's prescribed cloud droplet number concentration of 100 $cm^{-3}$ leads to much weaker precipitation. This study uses prognostic aerosol (*37*) and focuses on a case with low aerosol concentrations in the MBL (i.e., MBL_30) because it has potential for a strong marine cloud brightening (MCB) response (*38, 39*), so it helps understand how the best-case MCB scenario would change with climate, which will be further discussed later.

In the perturbed cases, a spatially-uniform increase in the surface aerosol flux before the first daytime (the pink-shaded areas in Figure 1) leads to perturbed aerosol concentrations Na (MBL-average) of approximately 130 $cm^{-3}$, about 4 times larger than in the unperturbed simulations (dashed vs. solid lines in Figure 1b). Observations (*40*) have shown that accumulation-mode aerosol concentrations can reach up to 150 cm$^{-3}$ in the subtropical Northeast Pacific, suggesting that even the elevated perturbed aerosol levels in our study are within a realistic range of present-day aerosol conditions.

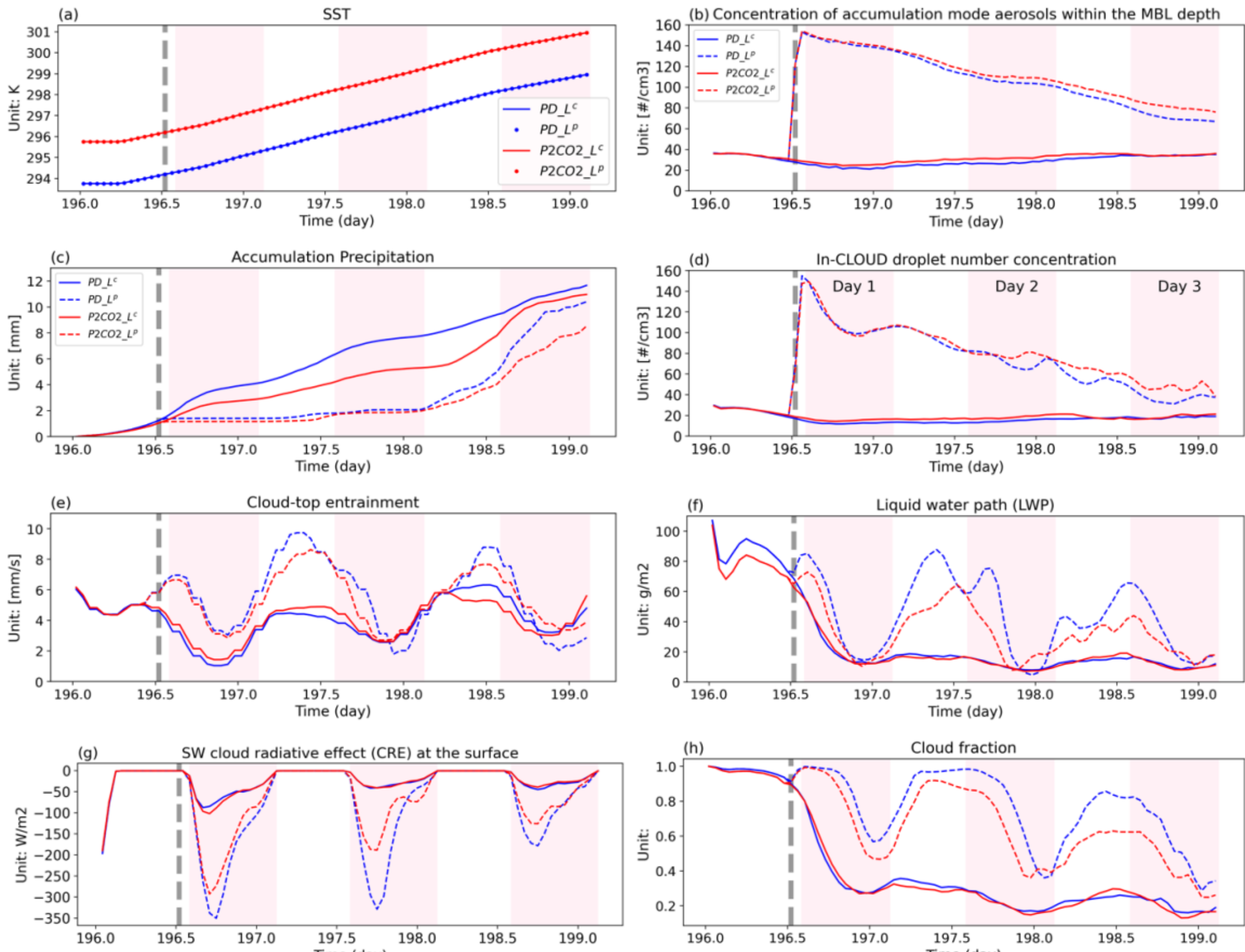

**Fig 1. Time series of cloud-related variables.** Time series of sea surface temperature (a), MBL-average accumulation-mode aerosol concentration (b), accumulation precipitation (c), in-cloud droplet number concentration (d), cloud-top entrainment rate (e), liquid water path (f), SW cloud radiative effect at the surface, and cloud fraction (h) with (dashed lines) and without (solid lines) aerosol perturbations in the present day (blue) and a doubled-$CO_2$ (red) climate, from the first four simulations listed in Table 1. The pink-shaded areas indicate the daytime periods in Day 1, Day 2, and Day 3. The dashed grey indicates when the aerosol perturbations are induced in the perturbed cases.

Aerosols induced in the perturbed cases can serve as CCN, forming more smaller cloud droplets through activation and condensation processes, which is known as the first indirect effect of aerosols (or Twomey effect (*41, 42*)). Compared to the control cases, the larger aerosol concentrations in the perturbed cases increase cloud droplet number concentration (Nc) to as high as 150 cm$^{-3}$ (Figure 1d). These more smaller droplets also inhibit precipitation formation and, in this case, increases the lifetime of overcast stratocumulus clouds (*43*) (Figure 1h). Subsequently, Nc gradually decreases in the 3-day simulation (Figure 1b), driven by cloud-top entrainment and scavenging (Figure S3).

By comparing aerosol-induced cloud changes between the present day and a doubled-$CO_2$ climate (e.g., blue vs. red lines in Figure 1), we can evaluate aerosol-cloud interactions in a changing climate. The aerosol-induced increase of Nc (Figure 1d) in a doubled-$CO_2$ climate is almost identical to that in the present day, especially on Day 1, when mainly stratocumulus clouds are present. Compared to the P2CO2 case, the PD case shows a slightly larger scavenging sink (Figure S3.b) on Day 3, resulting in a smaller Nc concentration (Figure 1d). But ΔNc caused by the activation process (Figure S2.b) does not change much in a changing climate, indicating that the Twomey effect associated with a given aerosol perturbation would not change much in a changing climate.

Since rain formation relies on larger cloud droplets, the more numerous and smaller cloud droplets in the perturbed cases (induced by aerosol perturbations) inhibit auto-conversion and accretion when compared to the control cases (Figure S3.b), which further depresses the precipitation (Figure 1c) and contributes to an increase in the liquid water path (Figure 1f) and cloud cover (or cloud fraction) (Figure 1h). These changes (or adjustments) in CF and liquid water path (LWP) represent the second indirect effect of aerosols (*43, 44*). The stratocumulus clouds, which cannot be maintained in the control cases, are, in contrast, formed and maintained in the perturbed cases (dashed lines in Figure 1h).

Figures 1f and 1h show that the aerosol-induced LWP and CF adjustments are weaker in the doubled-$CO_2$ climate (P2CO2_$L^p$ vs. P2CO2_$L^c$) than in the present day (PD_$L^p$ vs. PD_$L^c$). If we shift our perspective and consider the impact of a warmed climate by comparing the two cases with aerosol perturbations (P2CO2_$L^p$ vs. PD_$L^p$), we find cloud changes with warming that are consistent with other studies (*25, 45*), indicating that stratocumulus cloud formation favors the present day over a doubled-$CO_2$ climate. This is because global warming tends to thin and potentially break up stratocumulus clouds by several mechanisms, including the entrainment liquid-flux adjustment (*36, 46*).

**Marine boundary layer structure**

From Day 1 to Day 3, stratocumulus clouds tend to break up and transition to cumulus clouds, mainly driven by the increase of SST (*28, 34*). As previous studies indicate that the most salient feature of SCT is the structural change in the boundary layer circulation (*27, 34*), Figure 2 shows the daytime-mean cloud fraction (shaded) and boundary-layer structure (black, orange, and red bars) (*25*) for the three days. The black bar in Figure 2 is the inversion height Zi, where the relative humidity crosses 50%. The red bar is the lifting condensation level (LCL) of air with the properties at height z = 0.1×Zi, representing the lowest cloud base. The orange bar is the lifting condensation level (LCL) of air with properties at height z = 0.9×Zi, representing the stratocumulus cloud base. There is no stratocumulus cloud layer if the orange bar is near or above the black bar. The distance between the orange and red bars can be defined as a decoupling index:

$$\Delta LCL = LCL\ (z = 0.9\times Zi) - LCL\ (z = 0.1\times Zi)$$

The cloudy boundary layer is well-mixed if the orange and red bars are close, indicating a less decoupled boundary layer.

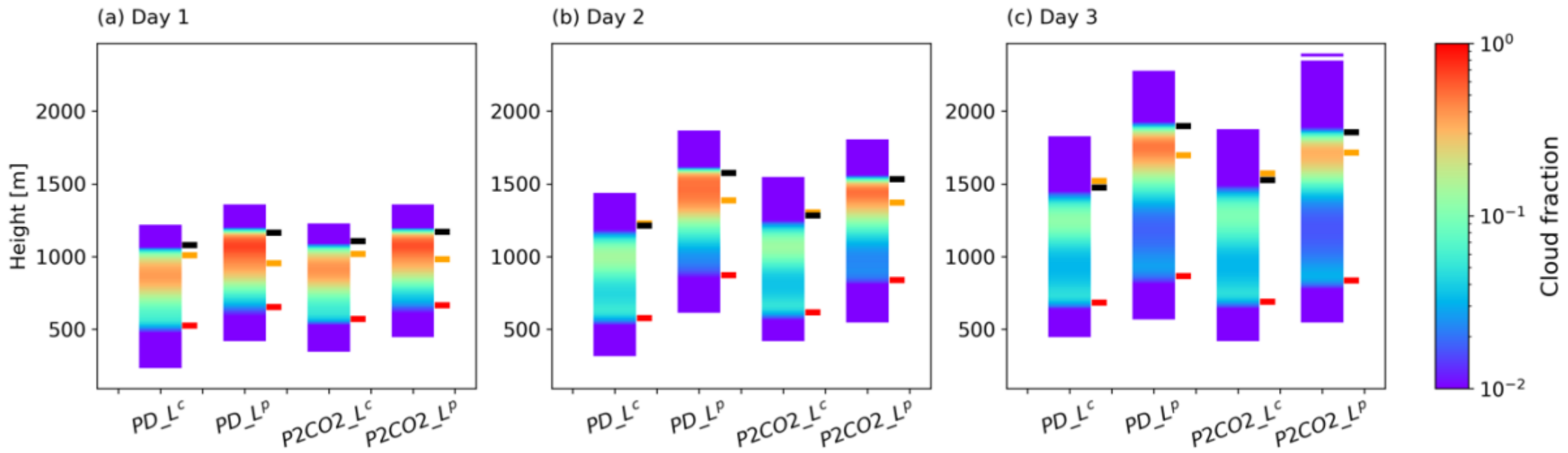

**Figure 2. Cloud and the marine boundary layer.** Vertical profile of cloud fraction (shaded) and boundary layer structure, where the black bar indicates inversion height (Zi), the orange bar the lifting condensation level (LCL) of air with properties at height z = 0.9×Zi, and the red bar the LCL of air with properties at height z = 0.1×Zi. All values are averaged over the whole domain and averaged over daytime (i.e., pink-shaded areas in Figure 1) on Day 1 (a), Day 2 (b), and Day 3 (c). The x-axis indicates the first four simulations listed in Table 1.

On Day 1, the perturbed cases show a stratocumulus cloud layer (between the black and orange bars in Figure 2a), while there are almost no stratocumulus clouds in the control cases because the low aerosol conditions lead to persistent precipitation and cloud breakup (*47*). The perturbed cases show a more well-mixed stratocumulus-topped boundary layer (compared to the control cases), as indicated by the smaller distance between the red and orange bars in Figure 2a. As shown in Figure S4, there are two turbulent kinetic energy (TKE) peaks (produced by buoyancy and wind shear) in the MBL: one in the cloud layer (dominated by buoyancy production) and another near the surface (dominated by shear production). The perturbed cases have more TKE in the cloud layer while the control cases have more TKE near the surface. Cloud-top entrainment (Figure 1e), driven by LW radiative cooling at the top of the stratocumulus clouds, is much larger in the perturbed cases than in the control cases.

In the control case without aerosol perturbations, the cloud layer shifts slightly upward (Figure S5) in the global warming scenario (in our model settings, subsidence is decreased in a warmer climate, as is expected with a weaker overturning tropical circulation (*48, 49*)). In the perturbed case with increased aerosols, global warming leads to a decrease in the height of the cloud layer, consistent with weaker cloud-top entrainment (Figure 1e) as more CO2 and water vapor in the free troposphere (FT) inhibit cloud-top longwave cooling (*46*).

From Day 1 to Day 3, clouds transition from stratocumulus to cumulus, with the deepening (Zi tends to increase) and decoupling (ΔLCL tends to increase) of the MBL (Figure 2). Compared to the present day ($PD_L^p$), global warming ($P2CO2_L^p$) decreases the thickness of stratocumulus clouds (i.e., the distance between black and orange bars in Figure 2) in the perturbed cases but doesn't change the boundary layer structure (e.g., decoupling index ΔLCL is similar in $PD_L^p$ vs. in $P2CO2_L^p$). As pointed out in previous studies (*25, 46*), global warming can cause thinning of the cloud layer even with little change in the buoyancy fluxes there.

**Cloud radiative effects**

Figure 1g shows that injected aerosols can notably strengthen shortwave (SW) cloud radiative effects (CRE) at the top of the atmosphere, causing a larger cooling due to aerosol-induced increases in cloud cover (Figures 1d, 1f, 1h). Because the CRE is the change in the downward radiative flux from clear-sky to full-sky conditions that include clouds, the negative SW CRE in Figure 1g corresponds to a cooling effect on the climate system relative to clear sky. Conversely, positive CRE implies a warming effect. Table 2 shows that the change of CRE (△CRE) induced by aerosols is weakened by >30% in a doubled-$CO_2$ climate (compared to the present day) in our simulations over the Northeast Pacific Ocean.

To further explore how increased aerosols change the cloud radiative effect, we first decompose the aerosol-induced △CRE into two components (See Methods), representing contributions from the aerosol-induced △CF (change of cloud fraction) and △A (change of cloud albedo):

$$\triangle CRE = \triangle CRE_CF + \triangle CRE_A$$

In both PD_L and P2CO2_L cases in Table 2, △CF has much larger contributions to the △CRE than the contribution from △A (i.e., △CRE_CF vs. △CRE_A). Aerosol-induced △CF and △A are smaller in the doubled-$CO_2$ climate than in the present day.

Furthermore, △CRE_A originates from aerosol-induced △Nc (change of in-cloud droplet number concentration) and △LWP (change of in-cloud liquid water path):

$$\triangle CRE_A = \triangle CRE_A_{Nc} + \triangle CRE_A_{LWP} + \triangle CRE_A_{Res}$$

Where $\triangle CRE_A_{Res}$ is a negligible residual term. $\triangle CRE_A_{Nc}$ in Table 2 shows a cooling due to the Twomey effect (aerosols induce more smaller cloud droplets that reflect more shortwave radiation), which is almost identical between the present day to the doubled-$CO_2$ climate. This result, that the Twomey effect is not significantly influenced by global warming, is consistent with the nearly identical changes in Nc found in Figure 2d. Aerosol-induced △LWP has two ways to contribute to the △CRE. On the one hand, $\triangle CRE_A_{LWP}$ (due to in-cloud LWP) in Table 2 shows a slightly warming radiative effect due to the aerosol-induced decrease of cloud thickness (Figure 2 and Figure S5), which is a warming contribution to △CRE_A. On the other hand, overall, there is an increase in aerosol-induced domain-average LWP (Figure 1f), consistent with the increase of CF (Figure 1h), which has a cooling contribution to △CRE_CF.

**Table 2. Aerosol-induced cloud radiative effects.**

Changes of the shortwave cloud radiative effect (△CRE, Unit: W $m^{-2}$) due to aerosol perturbations and the decomposed contributions from changes of cloud fraction (△CRE_CF) and cloud albedo (△CRE_A). The contribution from cloud albedo (△CRE_A) is further decomposed into the contributions from droplet number concentration (△CRE_$A_{Nc}$), and liquid water path (△CRE_$A_{LWP}$), with a negligible residual term (△CRE_$A_{Res}$). All values are 3-day averages.

| Simulations | △CRE | △CRE_CF | △CRE_A | △CRE_$A_{Nc}$ | △CRE_$A_{LWP}$ | △CRE_$A_{Res}$ |
|---|---|---|---|---|---|---|
| PD_$L^p$ - PD_$L^c$ | -48.0 | -40.3 | -7.8 | -8.9 | 2.1 | -1.0 |
| P2CO2_$L^p$ - P2CO2_$L^c$ | -33.1 | -27.3 | -5.8 | -8.5 | 3.0 | -0.3 |

Note: PD_$L^p$, PD_$L^c$, P2CO2_$L^p$, and P2CO2_$L^c$ are the first four simulations listed in Table 1.

**Sensitivity to aerosols in the unperturbed MBL (MBL_30 vs. MBL_100)**

To highlight aerosols' interactions with clouds in a case with strong potential for cloud thickening in response to natural or anthropogenic aerosol perturbations, the control case in this study has an initially low MBL-average aerosol concentration (Na) of 30 $cm^{-3}$ (i.e., MBL_30). To test the sensitivity of our results to different initial (i.e., unperturbed) MBL aerosol concentrations, we carried out a set of 1-day simulations using an initially higher MBL-average Na of 100 $cm^{-3}$ (i.e., MBL_100). These short simulations focus on how stratocumulus early in the Sc-to-Cu transition respond to aerosol perturbations under different background MBL aerosols (MBL_30 vs. MBL_100). In this new set of perturbed simulations in MBL_100 with the same increase of surface aerosol flux, we find that:

- Unlike the MBL_30 where the stratocumulus clouds are not maintained (solid blue line in Figure S6h) due to the high precipitation (solid blue line in Figure S6c), the higher initial MBL aerosol concentration in MBL_100 inhibits precipitation (solid yellow line in Figure S6c) and maintain full stratocumulus cloud cover (solid yellow line in Figure S6h), even before the aerosol perturbation is introduced.
- Unlike the MBL_30 where aerosol perturbation induced substantial increases in cloud cover and boundary layer structure (e.g., PD_$L^p$ vs. PD_$L^c$), the aerosol perturbation in MBL_100 causes a much weaker cloud response (e.g., PD_$H^p$ vs. PD_$H^c$) because the unperturbed MBL_100 simulation has more widespread cloud cover. When aerosol perturbations are introduced into the MBL_100, most aerosol-induced cloud changes (e.g., △CF, △LWP, △Nc) are much weaker on the first day, compared to simulations with MBL_30 (Table 3).
- Unlike the MBL_30 where the aerosol perturbation inhibits precipitation and increase stratocumulus cloud (△RWP, △LWP and △CF in Table 3), the aerosol perturbation has negligible influences on precipitation in MBL_100 and may even slightly decrease the cloud fraction and LWP of stratocumulus clouds (△LWP and △CF in Table 3) due to the increase of entrainment and evaporation at the top of the clouds.

The opposite responses of CF and LWP to aerosol perturbations in MBL_30 (increased CF and LWP due to aerosol-induced precipitation inhibition) versus in MBL_100 (decreased CF and LWP due to aerosol-induced increase of entrainment and evaporation) indicate high uncertainties in aerosol-induced CF and LWP adjustments driven by complex processes (*44, 50-52*).

**Table 3. Aerosol-induced cloud changes in MBL_30 vs. MBL_100.**

Aerosol-induced changes in the MBL-average accumulation-mode aerosol concentration (Na, unit: $cm^{-3}$), in-cloud droplet number concentration (Nc, unit: $m^{-3}$), rain water path (RWP, unit: g $m^{-2}$), liquid water path (LWP, unit: g $m^{-2}$), cloud fraction (CF), cloud-top entrainment (Entr, unit: mm $s^{-1}$), and shortwave cloud radiative effect (CRE, unit: W $m^{-2}$). All values are averaged over the daytime on Day 1 (pink-shaded areas in Figure 1). All simulations are described in Table 1.

| Initial MBL | Simulations | △Na | △Nc | △RWP | △LWP | △CF | △Entr | △CRE |
|---|---|---|---|---|---|---|---|---|
| MBL_30 | $PD_L^p$ - $PD_L^c$ | 120 | 102 | -5.7 | 18 | 0.41 | 2.88 | -125 |
| MBL_100 | $PD_H^p$ - $PD_H^c$ | 109 | 77 | -0.3 | -6 | -0.04 | -0.00 | -5 |
| MBL_30 | $P2CO2_L^p$ - $P2CO2_L^c$ | 119 | 99 | -4.2 | 10 | 0.32 | 2.12 | -90 |
| MBL_100 | $P2CO2_H^p$ - $P2CO2_H^c$ | 108 | 75 | -0.3 | -5 | -0.03 | 0.05 | -4 |

**Impacts of aerosols on low cloud feedbacks**

This study focuses on how aerosols influence clouds in a changing climate over the Northeast Pacific Ocean, where the low cloud feedbacks are dominated by changes in shortwave cloud radiative effects (SW CRE). From a different perspective, we may evaluate cloud feedbacks by comparing the SW CRE in the present day to those in a doubled- or quadrupled-$CO_2$ climate under different aerosol perturbations, as shown in Figure 3. SW CRE changes with doubling $CO_2$ (P2CO2 - PD) are weak when Nc is small (for Nc around 20 $cm^{-3}$), but SW CRE weakens by ~20 W $m^{-2}$ with a doubling of $CO_2$ when Nc is larger (for Nc around 110 $cm^{-3}$). The change of SW CRE caused by quadrupling $CO_2$ (~40 W $m^{-2}$) is about twice as large as that by doubling $CO_2$, especially when Nc>100 $cm^{-3}$ (Figure 3). The stars in Figure 3 represent the SW CRE in the present day ($PD$) and in a 4×$CO_2$ climate (P4CO2), as given in BB2014 using simulations with fixed Nc of 100 $cm^{-3}$. The slight difference in SW CRE and cloud feedbacks between our results and BB2014 is expected due to changes in model settings: our simulations predict rather than prescribe aerosol and cloud droplet concentrations, and we have a much larger horizontal domain (51.2 km × 51.2 km) than does BB2014 (4.48 km × 4.48 km).

Comparing the two control cases (P2CO2_L$^c$ vs. PD_L$^c$) suggests that low cloud feedbacks might be weaker under very low aerosol conditions, mainly because overcast stratocumulus clouds cannot be maintained when low aerosol concentrations promote precipitation. Simulations with global climate models have found similar results: aerosol forcing and cloud feedbacks are related through cloud processes and depend on the mean state of clouds (*9*).

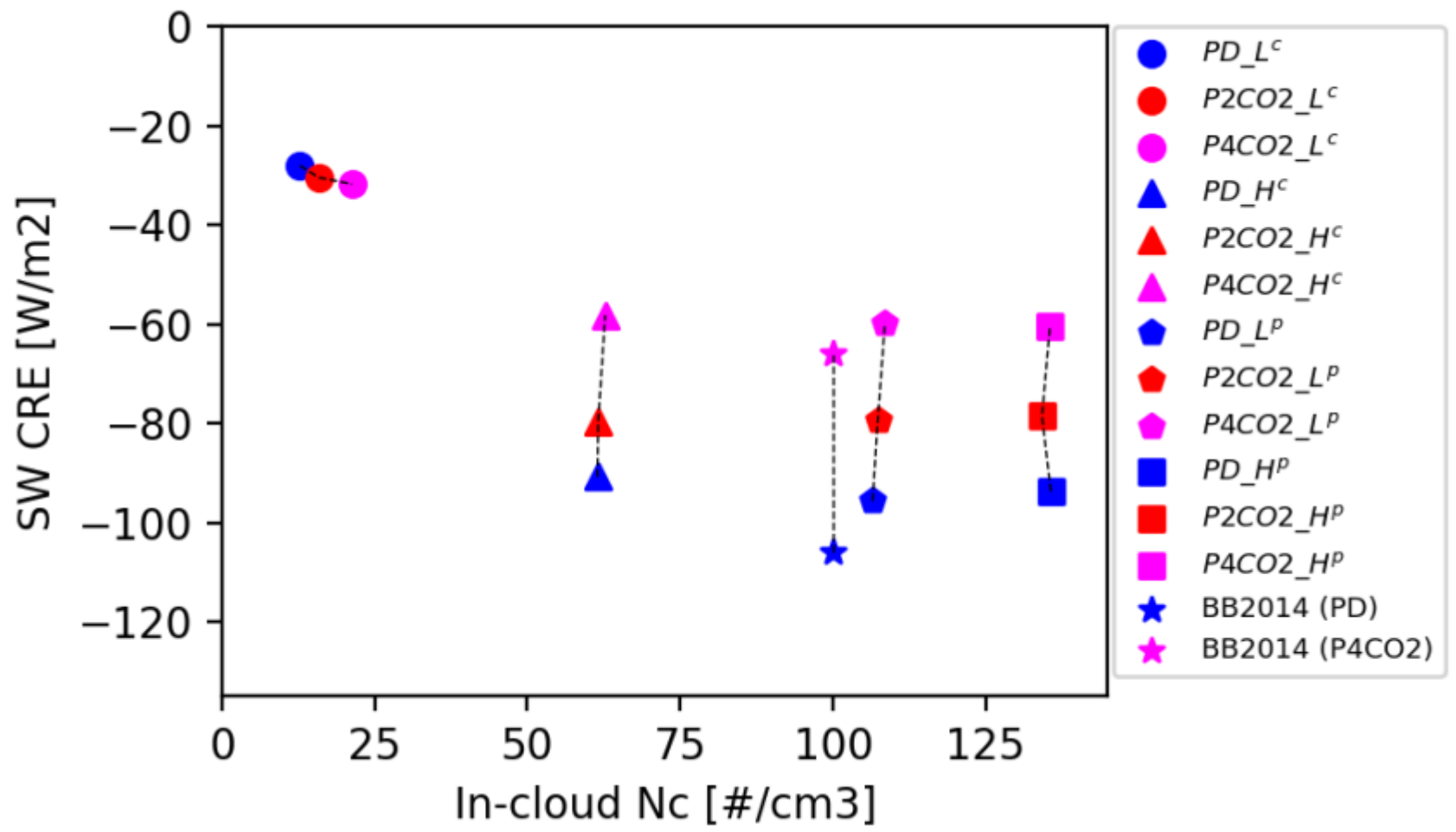

**Figure 3. Shortwave (SW) cloud radiative effect (CRE) versus in-cloud droplet number concentration (Nc).** Domain-average SW CRE vs. in-cloud Nc averaged on Day 1 (including daytime and nighttime). Each data point represents a simulation listed in Table 1, except the two stars that represent simulations from BB2014 in the present day (PD) and a $4\times CO_2$ climate (P4CO2).

**Implications for Marine Cloud Brightening (MCB)**

As an application of aerosol-cloud interactions, marine cloud brightening (MCB) describes the idea of intentionally injecting aerosols into stratocumulus clouds, the most common cloud type covering more than 20% of the ocean surface (*53, 54*). These additional aerosols would lead to more numerous and smaller cloud droplets, which can increase cloud albedo and reflect more incoming solar radiation (*55-58*). Our results indicate that global warming can weaken the cooling effects of MCB ($\Delta CRE$ in Table 2) mainly by inhibiting the second indirect effect of aerosols (Figure 4).

This study also indicates that the cooling effect of MCB is weaker in an MBL with higher background aerosol concentrations (MBL_100 vs. MBL_30). For example, implementing MCB in the conditions given by MBL_100 may lead to a decrease in cloud thickness (e.g., negative $\triangle$LWP shown by triangles in Figure 4b) due to aerosol-induced increase in entrainment and evaporation, which will have a negative contribution to MCB's cooling effects. It is worth mentioning that anthropogenic aerosols are currently declining in many locations, and in some locations rapidly (*59*). Such efforts may lead to more widespread regions of broken and strongly precipitating marine low clouds in the future, a potential example of “aerosol-mediated cloud feedbacks” (*60*). However, those clouds would also be more amenable to MCB (*61*).

Our results can help with the design of MCB injection strategies to achieve a stronger cooling effect (e.g., applying MCB preferentially to areas of the MBL with lower aerosol concentrations, or by adjusting the MCB injection rate based on background aerosol concentration). Figure 3 shows that increasing Nc leads to much stronger cloud brightening (i.e., more negative SW CRE changes) before reaching a saturation (or ceiling). This MCB’s saturation (ceiling) cooling effect is approximately -100 W $m^{-2}$ (averaged on the 1st day) in PD conditions, and decreases to -80 W $m^{-2}$ in P2CO2 and -60 W $m^{-2}$ in P4CO2 conditions, in our study.

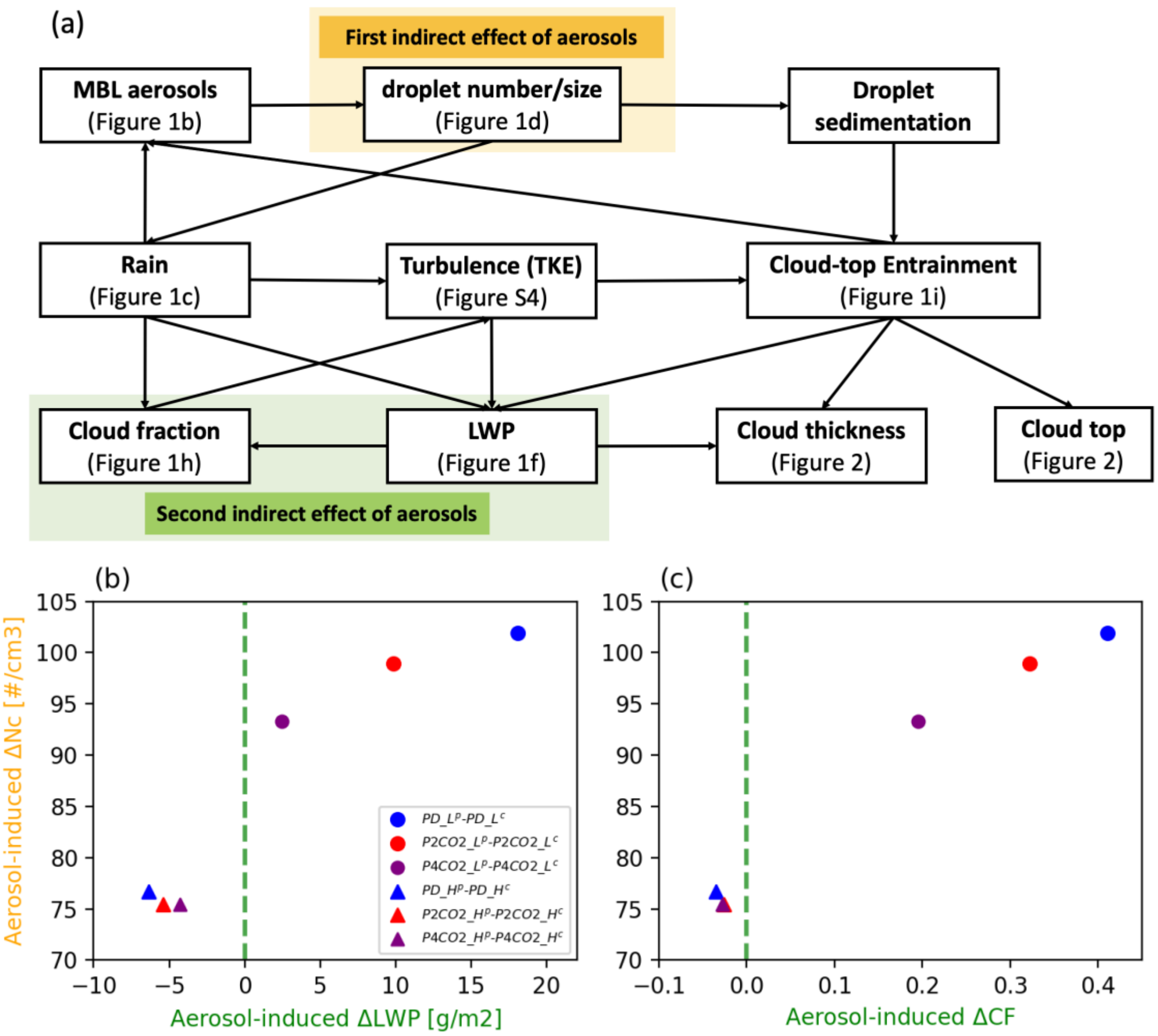

**Figure 4. Aerosol-cloud interactions, including first (Twomey effect) and second (adjustment of cloud fraction and liquid water path) indirect effects of aerosols.** (a) Schematic of aerosol-cloud interactions. (b) Aerosol-induced change in cloud droplet number concentration (△Nc) versus the change in liquid water path (△LWP) between perturbed and control simulations (listed in Table 1) averaged over the first daytime (i.e., pink-shaded areas in Figure 1). (c) △Nc versus Aerosol-induced change of cloud fraction (△CF), similar to (b).

## Discussion

Through inducing a temporary increase in the surface aerosol flux, we study how increased aerosols influence clouds (Figure 4a) in the NE Pacific marine boundary layer in a changing climate (i.e., from PD to P2CO2). Figure 4a summarizes the aerosol-cloud interactions discussed in this study, which employs high-resolution LES of the SCT transition along quasi-Lagrangian (airmass-following) trajectories. The effects of changes in aerosols and climate on such SCT simulations have been studied before (*32-34, 36, 47*), but few studies have characterized the interactions of aerosol and climate change in such simulations. Compared to the present day, the doubled-$CO_2$ climate slightly weakens the aerosol-induced △Nc (by less than 5%) but leads to stronger decreases in the aerosol-induced △LWP (>40%, red vs. blue circles in Figure 4b) and the aerosol-induced △CF (>20%, Figure 4c). Thus, global warming slightly modulates the Twomey effect (aerosol-induced △Nc) but notably inhibits the second indirect effects of aerosols (aerosol-induced △CF and △LWP), which results in an inhibition of aerosol-cloud interactions in a doubled-$CO_2$ climate.

To explore how aerosol-induced cloud changes influence the cloud radiative effect, we decompose the aerosol-induced changes in cloud radiation effects (△CRE) into contributions from the aerosol-induced △CF, △LWP, and △Nc (Table 2). From the control case with low background aerosol (MBL_30) to the perturbed case (increased aerosols shown in Figure 1b), aerosol-induced △CF shows the largest contribution (cooling effects) to the △CRE. Aerosol-induced △Nc causes a cooling effect by increasing cloud albedo due to the Twomey effect. Aerosol-induced changes to cloud macrophysical properties include a cooling contribution to △CRE (related to the increase of cloud fraction) and a warming contribution to △CRE (related to the decrease of cloud thickness). All analysis in this paper focuses on daytime (shortwave radiative effects), a comparison between daytime and nighttime (diurnal evolution of aerosol-cloud interactions in a changing climate) will be our next step.

This study has explored the interplay between aerosol-cloud interactions (the most uncertain climate forcing) and cloud responses to climate change (i.e., cloud feedbacks), finding that aerosol-cloud interactions weaken in a warmer climate in these large eddy simulations of marine low cloud transitions. While the forcing-feedback relationships in an ensemble of global models have been explored (*9*), we have shed light on these relationships in marine low clouds during the transition from stratocumulus in simulations that explicitly represent cloud, aerosol, precipitation, and turbulence variations that are parameterized in larger-scale models. We find that potential cooling by marine cloud brightening in a warmer climate is less than under present-day conditions, but that changes in aerosol emissions that occur alongside warming could modulate that potential.

In this study, we focus on how aerosol-cloud interactions respond to climate change, combining all warming perturbations (Figure S1). In the next step, we will evaluate how each component of composite warming perturbations (e.g., the increase of SST, the decrease of subsidence) influences ACI, which can help us better understand the performance of cloud controlling factors (CCFs) in a changing climate. In the clean case, the increase of aerosols (due to less precipitation caused by global warming) is the dominant CCF that facilitates cloud formation, while in the polluted case with enough aerosols, the decrease of turbulence in a doubled-$CO_2$ climate (due to reduced longwave radiative cooling) thins and lowers clouds (seen in the distance between and height of the black and orange bars in Figure 2).

## Methods

### SAM model settings.

We used the SAM model to carry out the large-eddy simulations along a Lagrangian trajectory following a previous study referred to here as BB2014 (*36*) with the domain center starting at (25N, 125W) in the NE Pacific. The SAM model has 288 vertical levels with a model top at 4.2 km. The atmosphere above the model top is included in computations of radiative heating, based on a sounding that is included in the model forcings. The horizontal domain size is 51.2 km × 51.2 km, with a horizontal resolution of 100 m. The total simulation time is 3 days in summertime (day 196.1-199.1). The first half day (day 196.1-196.6) is considered the spin-up time. The transient nudging of temperature (T), water vapor (q), and aerosols is applied in the marine boundary layer (MBL) in the first 6 hours. T and q are nudged in the free troposphere (FT) in the whole 3-day simulation.

The SAM model includes an interactive aerosol scheme (*37*) with representation of the PBL aerosol budget for a single accumulation mode, with sources and sinks from the surface and FT. Before the aerosol perturbation, the initial background aerosol concentration is 70 $cm^{-3}$ in FT (*36, 40*) and 30 $cm^{-3}$ in the MBL (i.e., MBL_30).

### Global warming settings.

Compared to PD, P2CO2 represents a doubled-$CO_2$ climate (*36*). For P2CO2 (compared to PD), the $CO_2$ concentration is increased to 2 times (2×$CO_2$), SST and absolute temperature are increased by 2 K, subsidence is reduced by ~5%, and specific humidity is increased to keep the relative humidity identical to that in PD. More details can be found in Figure S1.

### Calculation of cloud-top entrainment.

In this study, cloud-top entrainment ($\omega_e$) is calculated as:

$$\omega_e = \frac{dZ_{inv}}{dt} - \omega_{ls,inv}$$

Where $\frac{dZ_{inv}}{dt}$ is the rate of change of inversion height ($Z_{inv}$) over time, $\omega_{ls,inv}$ is the large-scale vertical velocity at the inversion height.

**Decomposing aerosol-induced ΔCRE into contributions from cloud albedo and CF.** For two simulated cases, "control" and "perturbed" (Table 1), the change in shortwave cloud radiative effects between the two cases (dashed vs. solid lines in Figure 1g) can be calculated as:

$$\Delta\text{CRE} = \text{CRE}^{\text{p}} - \text{CRE}^{c} \;\ldots\ldots\; (1)$$

$\text{CRE}^{\text{p}}$ is the cloud radiative effect (CRE) from the perturbed case (e.g., $PD^p$ and $P2CO2^p$ in this study), $\text{CRE}^c$ is the CRE from the control case ($PD^c$ and $P2CO2^c$ in this study), and ΔCRE is the change in CRE between the two cases (e.g., $PD^p$ vs. $PD^c$). CRE (always negative) can be calculated as:

$$\text{CRE} = \text{A} \cdot \text{S} - \text{S} \;\ldots\ldots\; (2)$$

Where S is the incoming shortwave (SW) flux at the top of the atmosphere (TOA), i.e. the clear-sky SW radiation. A · S represents the all-sky SW radiation. A is the total overcast albedo (i.e., the albedo as seen from space, also known as planetary albedo or TOA albedo) (*62-64*), which is calculated as:

$$\text{A} = \text{CF} \cdot A_{cld} + (1 - \text{CF}) \cdot A_{clr} \;\ldots\ldots\; (3)$$

CF is the cloud fraction, $A_{cld}$ is the overcast albedo over cloudy areas, $A_{clr}$ is the overcast albedo over the clear-sky areas. Based on Eq. (3), we can calculate the total overcast albedo for the control and perturbed cases:

$$\text{For control case: } A^c = CF^c \cdot A^c_{cld} + (1 - CF^c) \cdot A^c_{clr} \;\ldots\ldots\; (4)$$

$$\text{For perturbed case: } A^p = CF^p \cdot A^p_{cld} + (1 - CF^p) \cdot A^p_{clr} \;\ldots\ldots\; (5)$$

By combining Eqs. (1) and (2), we get:

$$\Delta\text{CRE} = \text{CRE}^{\text{p}} - \text{CRE}^{c} = (A^p \cdot \text{S} - \text{S}) - (A^c \cdot \text{S} - \text{S}) = S \cdot (A^p - A^c) \;\ldots\ldots\; (6)$$

In this study, we ensure that positive values of ΔCRE always mean warming effects. Substituting Eqs. (4) and (5) into Eq. (6):

$$\begin{aligned}\Delta\text{CRE} &= S \cdot \left[\left(CF^p \cdot A^p_{cld} + (1 - CF^p) \cdot A^p_{clr}\right) - \left(CF^c \cdot A^c_{cld} + (1 - CF^c) \cdot A^c_{clr}\right)\right] \\ &= S \cdot \left[CF^c \cdot \left(A^p_{cld} - A^c_{cld}\right) + (CF^p - CF^c) \cdot \left(A^p_{cld} - A_{clr}\right)\right] \;\ldots\ldots\; (7)\end{aligned}$$

Here, $S \cdot CF^c \cdot \left(A^p_{cld} - A^c_{cld}\right)$ can be considered as the contribution from the change of cloud albedo between two cases. $S \cdot (CF^p - CF^c) \cdot (A^p_{cld} - A_{clr})]$ can be considered as the contribution from the change of cloud fraction. Note that: $A_{clr} \approx A^c_{clr} \approx A^p_{clr}$.

**Decomposing the aerosol-induced ΔA into contributions from Nc and LWP.**
So far, we have decomposed the change of CRE between two cases into two parts: the first part represents the contribution from the change of cloud albedo ($\Delta A_{cld} = A^p_{cld} - A^c_{cld}$), the second part represents the contribution from the change of cloud fraction ($CF^p - CF^c$). In the next step, we will analyze how cloud number concentration (Nc) and liquid water path (LWP) inside clouds influence the change of cloud albedo. Based on previous studies (*35, 39, 64*), the changes in cloud albedo ($\Delta A_{cld}$) can be attributed in changes in Nc and LWP.

$$\Delta A_{cld} \approx \frac{\partial A_{cld}}{\partial \alpha_{cld}} \cdot \Delta\alpha_{Nc} + \frac{\partial A_{cld}}{\partial \alpha_{cld}} \cdot \Delta\alpha_{LWP} = \Delta A_{cld,Nc} + \Delta A_{cld,LWP} \text{ ...... (8)}$$

As mentioned before, the albedo (A) we have discussed so far is the overcast albedo (i.e., the albedo as seen from space, also known as planetary albedo or TOA albedo) (*62-64*). Using $\alpha$ as the albedo of the individual constituent, we can get:

$$A_{cld} = \alpha_{FT} + \alpha_{cld} \cdot \frac{T^2_{FT}}{1-\alpha_{FT}\cdot\alpha_{cld}} \text{ ...... (9)}$$

$$A_{clr} = \alpha_{atm} + \alpha_{sfc} \cdot \frac{T^2_{atm}}{1-\alpha_{atm}\cdot\alpha_{sfc}} \text{ ...... (10)}$$

$A_{cld}$ is the overcast albedo over the cloudy areas (seen from space), while $\alpha_{cld}$ is the cloud albedo. $\alpha_{FT}$ and $T_{FT}$ are the albedo and transmissivity of the free troposphere, respectively. $A_{clr}$ is the overcast albedo over clear-sky areas. $\alpha_{atm}$ and $T_{atm}$ are the albedo and transmissivity of the atmosphere, respectively. $\alpha_{sfc}$ is the surface albedo (i.e., the ratio of surface upward SW radiation to the surface downward SW radiation).

Thus, the change of overcast albedo over cloudy areas ($\Delta A_{cld}$) can be converted from the change of cloud albedo ($\Delta\alpha_{cld}$):

$$\Delta A_{cld,Nc} = \frac{\partial A_{cld}}{\partial \alpha_{cld}} \cdot \Delta\alpha_{Nc} = \frac{T^2_{FT}}{(1-\alpha_{FT}\cdot\alpha_{cld})^2} \cdot \Delta\alpha_{cld,Nc} \text{ ...... (11)}$$

$$\Delta A_{cld,LWP} = \frac{\partial A_{cld}}{\partial \alpha_{cld}} \cdot \Delta\alpha_{LWP} = \frac{T^2_{FT}}{(1-\alpha_{FT}\cdot\alpha_{cld})^2} \cdot \Delta\alpha_{cld,LWP} \text{ ...... (12)}$$

$\frac{\partial A_{cld}}{\partial \alpha_{cld}} = \frac{T_{FT}^2}{(1-\alpha_{FT}\cdot\alpha_{cld})^2}$ is derived from Eq. (9). To calculate $\Delta A_{cld,Nc}$ and $\Delta A_{cld,LWP}$ based on Eqs. (11) and (12), we need to calculate $\Delta\alpha_{cld,Nc}$, $\Delta\alpha_{cld,LWP}$, $\alpha_{cld}$, $T_{FT}$, and $\alpha_{FT}$, which is shown in the following.

The change of cloud albedo between two cases due to the change of Nc can be calculated as (*65*):

$$\Delta\alpha_{cld,Nc} = \frac{\alpha_{cld}^{c}\cdot(1-\alpha_{cld}^{c})(R_{Nc}^{\frac{1}{3}}-1)}{1+\alpha_{cld}^{c}\cdot(R_{Nc}^{\frac{1}{3}}-1)} \;\ldots\ldots\; (13)$$

The change of cloud albedo between two cases due to the change of LWP can be calculated as (*66*):

$$\Delta\alpha_{cld,LWP} = \frac{\alpha_{cld}^{c}\cdot(1-\alpha_{cld}^{c})(R_{LWP}^{\frac{5}{6}}-1)}{1+\alpha_{cld}^{c}\cdot(R_{LWP}^{\frac{5}{6}}-1)} \;\ldots\ldots\; (14)$$

$R_{Nc} = \frac{Nc^{p}}{Nc^{c}}$ is the ratio of the Nc in perturbed vs. control clouds. $R_{LWP} = \frac{LWP^{p}}{LWP^{c}}$ is the ratio of LWP in perturbed vs. control clouds. Note: both Nc and LWP should use the in-cloud values instead of the domain-average values from model results, in-cloud values can be derived by using model domain-average values and cloud fraction.

The cloud albedo can be derived from Eq. (9):

$$\alpha_{cld} = \frac{A_{cld}-\alpha_{FT}}{T_{FT}^{2}-\alpha_{FT}\cdot A_{cld}-\alpha_{FT}^{2}} \;\ldots\ldots\; (15)$$

$\alpha_{cld}^{c}$ used in Eqs. (13) and (14) is the cloud albedo from the control case, while $\alpha_{cld}$ used in Eqs. (11) and (12) is set as the average of cloud albedos from the control and perturbed cases (i.e., $\alpha_{cld} = \frac{\alpha_{cld}^{c}+\alpha_{cld}^{p}}{2}$, $\alpha_{cld}^{c}$ and $\alpha_{cld}^{p}$ can be calculated based on Eq. (15)). Note: the value assigned to $\alpha_{cld}$ can introduce bias for the final results, but the bias is negligible.

The transmissivity of the free troposphere ($T_{FT}$) is calculated as the ratio of downward SW radiation near the inversion height to the solar irradiation ($S$).

The albedo of the free troposphere can be calculated by:

$$\alpha_{FT} = \alpha_{atm}\frac{1-T_{FT}}{1-T_{atm}} \;\ldots\ldots\; (16)$$

Where $\alpha_{atm}$ is the albedo of the atmosphere, which can be estimated from Eq. (10). $T_{atm}$ is the transmissivity of the atmosphere (i.e., the ratio of downward SW radiation reaching the surface to the solar irradiation $S$ under the clear-sky condition).

## Acknowledgments

This work used Bridges-2 at Pittsburgh Supercomputing Center through allocation EES210037 from the Advanced Cyberinfrastructure Coordination Ecosystem: Services & Support (ACCESS) program, which is supported by National Science Foundation grants #2138259, #2138286, #2138307, #2137603, and #2138296.

**Funding:**

This research has been supported by the National Oceanic and Atmospheric Administration (grant nos. NA22OAR4310474 and NA20OAR4320271).

**Author contributions:**

Conceptualization: HS, RW, PB, SD
Methodology: HS, PB, EE, JYC
Investigation: HS, PB, RW
Software: HS, PB, EE, JYC
Visualization: HS, PB
Supervision: RW, PB
Writing—original draft: HS
Writing—review & editing: HS, PB, RW, EE, SD, JYC

**Competing interests:** The Authors declare that they have no competing interests.

**Data and materials availability:** All data, code, and materials used in the analysis will be uploaded and made available on Zenodo upon acceptance of the manuscript.

## Supplementary Materials

Figures S1 to S6.

# Supplementary Materials for

## Cloud-aerosol interactions in subtropical marine stratocumulus weaken in a warmer climate.

Hongwei Sun* *et al.*

*Corresponding author. Email: hongwei8@hawaii.edu

**This PDF file includes:**

Figures. S1 to S6

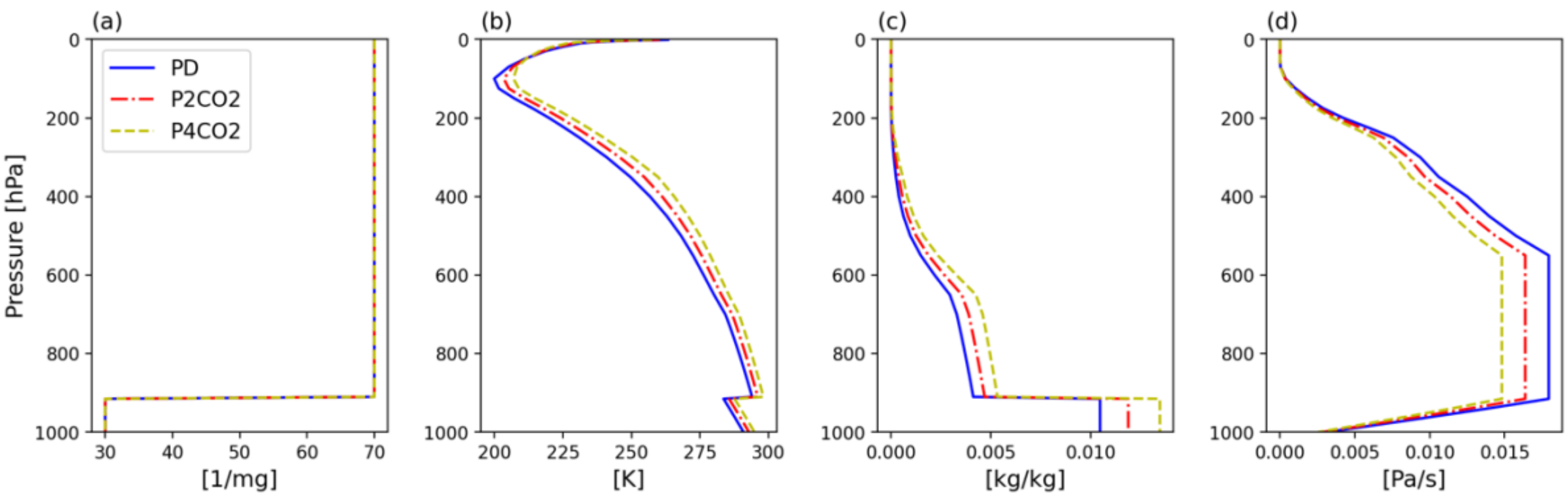

**Figure S1.**
Initial vertical profile of (a) accumulation mode aerosol concentration, (b) absolute temperature, (c) total specific humidity (vapor + liquid), and (d) large-scale vertical pressure velocity for simulations under present day (PD), doubled-$CO_2$ climate (P2CO2), and the quadrupled-$CO_2$ climate (P4CO2).

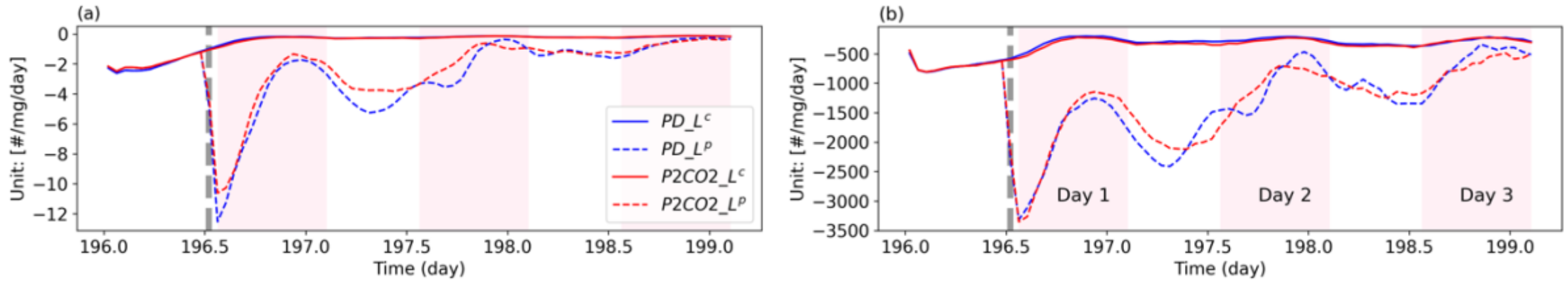

**Figure S2.**
Marine boundary-layer (MBL) averaged dry aerosol budget tendency due to (a) interstitial scavenging (caused by clouds and rain) and (b) activation. The results are from the first four simulations listed in Table 1. The pink-shaded areas indicate the daytime for Day 1, Day 2, and Day 3. The dashed grey indicates when the dry aerosol perturbations happen in the perturbed cases.

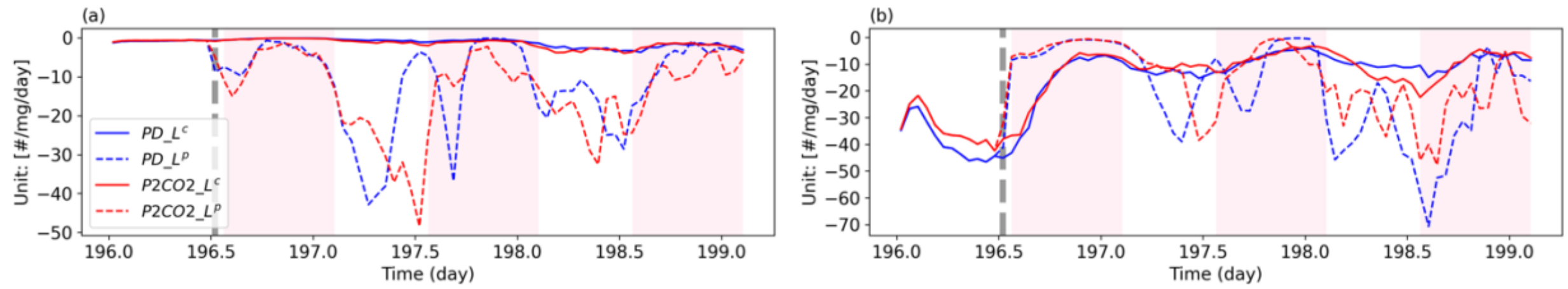

**Figure S3.**
MBL-average cloud droplet (Nc) budget tendency due to (a) entrainment and (b) scavenging (including auto-conversion and accretion).

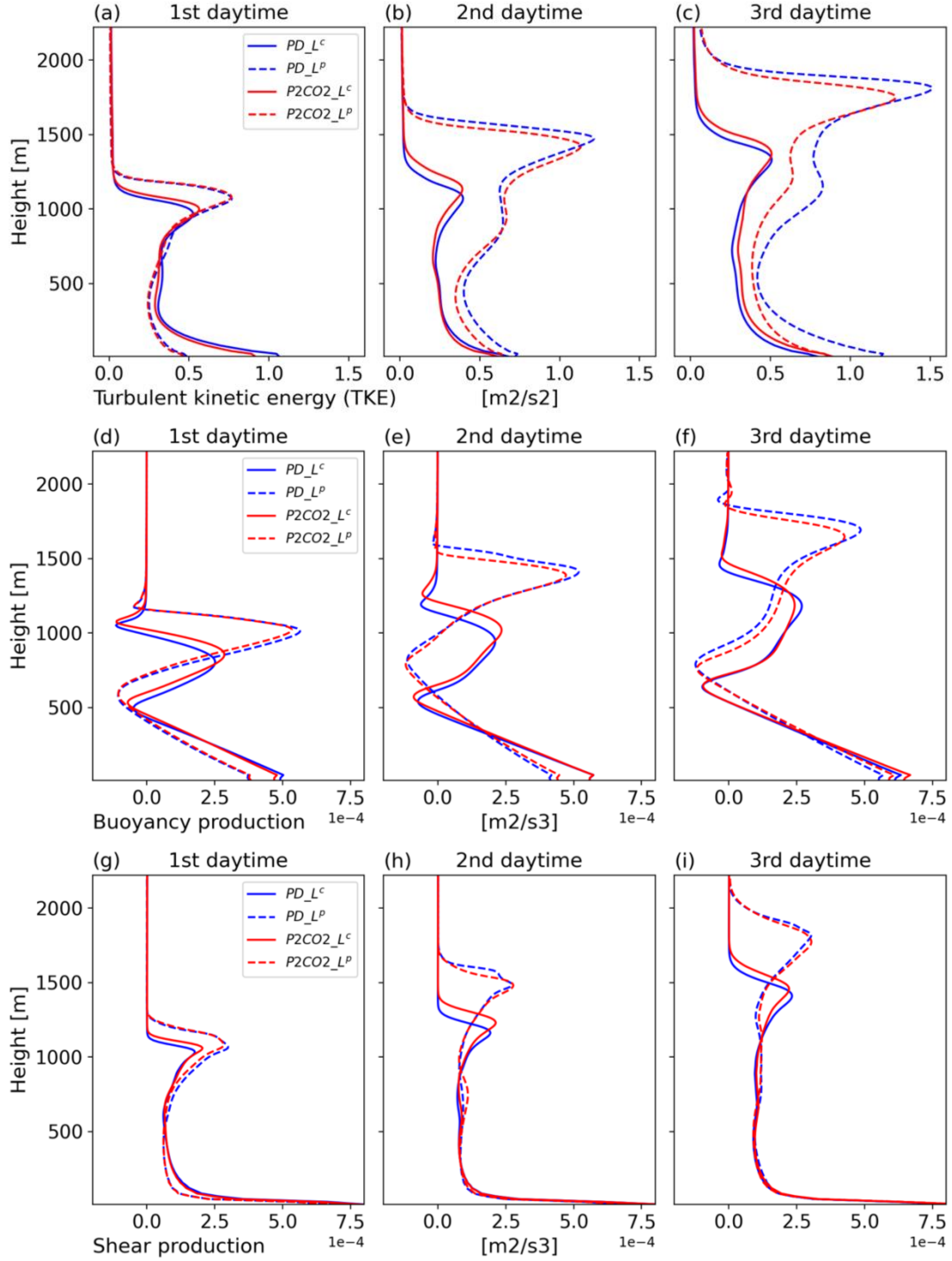

**Figure S4.**
Daytime-average vertical profiles of (a-c) turbulent kinetic energy (TKE), (d-f) buoyancy and (g-i) shear product of TKE from Day 1 to Day 3.

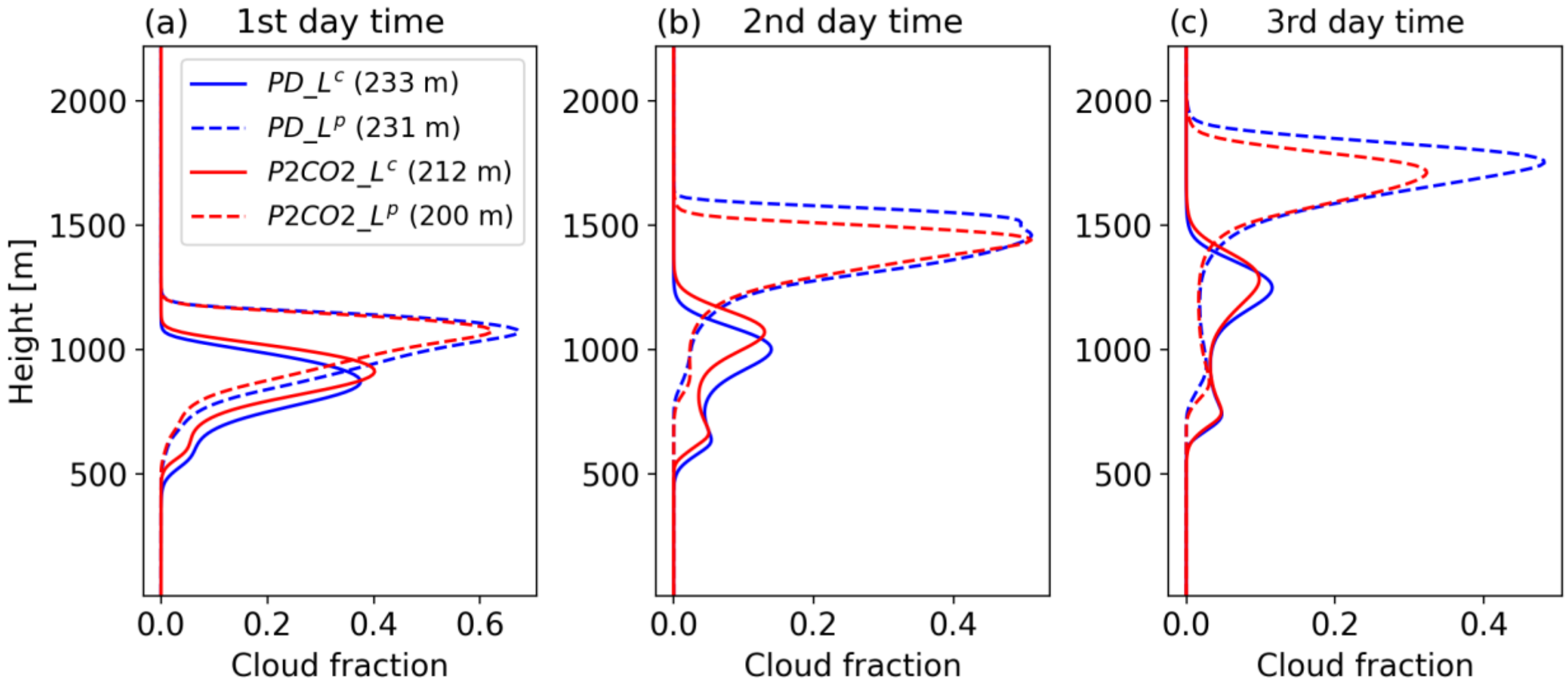

**Figure S5.**
Vertical profiles of cloud fraction and TKE for PD, P2CO2, and P4CO2. 1[st] daytime averaged cloud thickness is shown in the legend of Figure S5a.

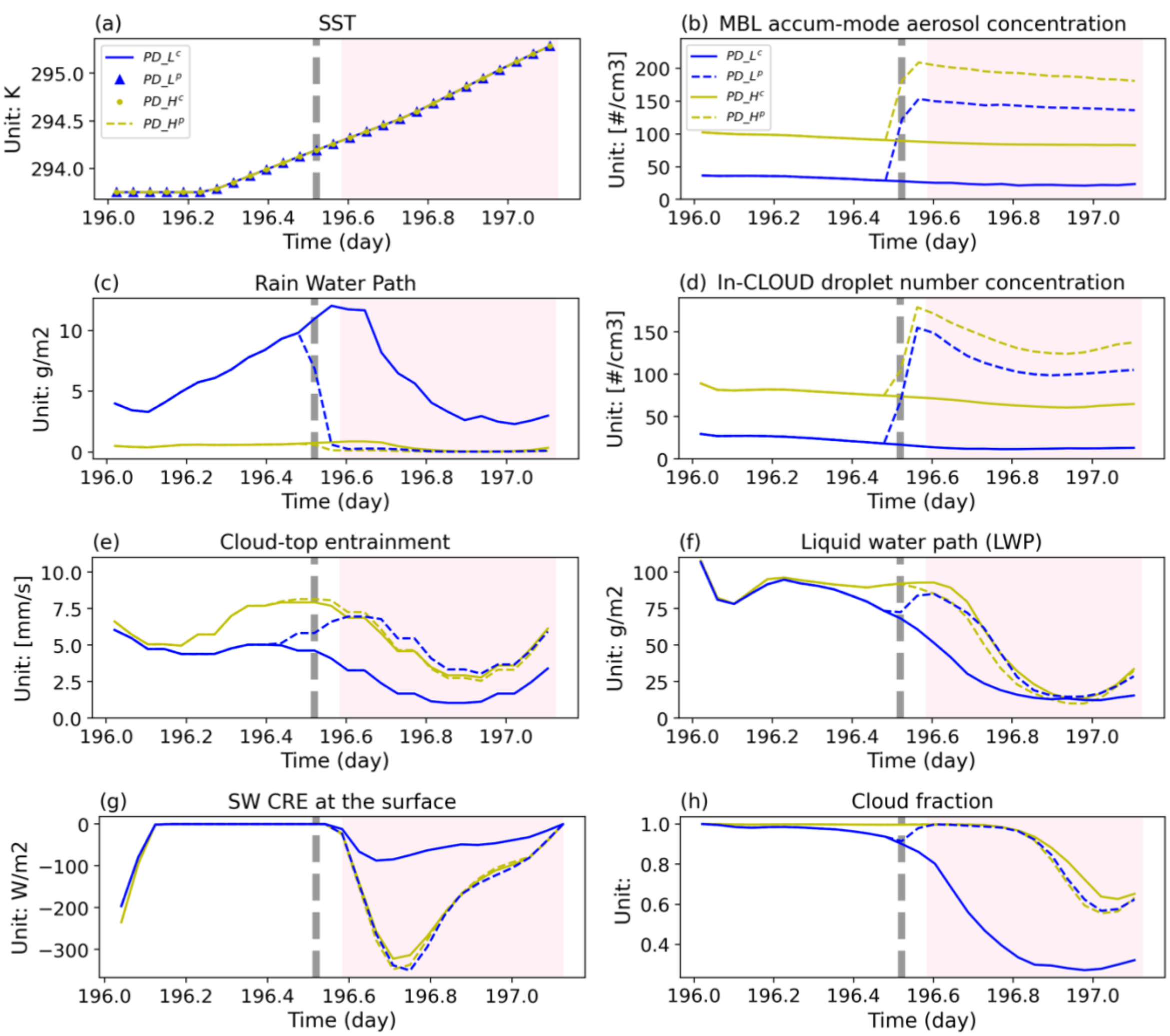

**Figure S6.**
Time series of different variables for clean MBL (yellow lines) vs. polluted MBL (red lines) in the present day (PD) with (dashed lines) vs. without (solid lines) aerosol perturbations.